# Frequency-independent voltage amplitude across a tunnel junction

Simon Feigl,[1,a)] Radovan Vranik,[1] Bareld Wit,[1] and Stefan Müllegger[1,2]

AFFILIATIONS
[1] Institute of Semiconductor and Solid State Physics, Johannes Kepler University Linz, 4040 Linz, Austria
[2] Linz Institute of Technology, Johannes Kepler University Linz, 4040 Linz, Austria

[a)] Author to whom correspondence should be addressed: simon.feigl@jku.at

## ABSTRACT

Radio-frequency (rf) scanning tunneling microscopy has recently been advanced to methods such as single-atom spin resonance. Such methods benefit from a frequency-independent rf voltage amplitude across the tunnel junction, which is challenging to achieve due to the strong frequency dependence of the rf attenuation in a transmission line. Two calibration methods for the rf amplitude have been reported to date. In this Note, we present an alternative method to achieve a frequency-independent rf voltage amplitude across the tunnel junction and show the results of this calibration. The presented procedure is applicable to devices that can deliver rf voltage to a tunnel junction.

Scanning tunneling microscopy (STM) methods have been in constant development since its invention around 40 years ago.[1,2] One recently added method is radio-frequency scanning tunneling spectroscopy: voltage changes with radio frequency (rf) are applied across the STM tunnel junction, while the differential tunneling conductance $(dI/dV)$ is recorded.[3] With this technique, we have demonstrated the excitation of electron and nuclear spin transitions in single molecules[4,5] and have revealed mechanical eigenmodes of molecular resonators.[6] A similar approach, well known as ESR-STM,[7–12] is based on single-spin magnetoresistance detection and has recently enabled measurement of single-atom spin resonance.

The frequency dependence of the rf voltage amplitude across the tunnel junction ($V_{pk,jun}$) may result in spurious measurement signals, which can easily be misinterpreted.[13,14] Therefore, a method to achieve constant $V_{pk,jun}$ independent of frequency is imperative. Currently, no method exists to measure $V_{pk,jun}$ directly,[11] but two approaches for the determination of $V_{pk,jun}$ and its calibration to a constant value have been reported.[13,14]

Compared to the method presented by Hervé et al.,[13] we detect the response of the tunnel junction to changes in the rf power in a simplified way. We measure the differential tunneling conductance for each frequency at one specific bias voltage instead of comparing conductance spectra across a wide bias voltage range. This spares the recording and comparison of additional conductance spectra and thus simplifies the calibration of $V_{pk,jun}$. Paul et al.[14] derived an analytical expression for $V_{pk,jun}$ as a function of the lock-in amplifier output signal, requiring an additional measurement and data fit. Our method spares these tasks; we adjust the rf voltage amplitude directly in response to the lock-in amplifier signal via a feedback loop. The calibration method presented in this Note is advantageous in cases where a small number of frequency values are needed as well as in cases where changes in the transfer function[15] occur frequently.

We have conducted the experiments with a modified Createc low-temperature STM under ultra-high vacuum (UHV) conditions. The sample temperature is typically 8 K, and the pressure is below $5 \times 10^{-11}$ mbar. We have upgraded the STM with rf-rated components, similar to our previously reported setup[4,6,16] (see Fig. 1). This setup enables rapid modulation of the voltage across the tunnel junction at high frequencies. The total voltage $V_{tot}$ is composed of two components, $V_{dc}$ and $V_{rf}$, which are superimposed,

$$V_{tot}(t) = V_{dc}(t) + V_{rf}(t). \qquad (1)$$

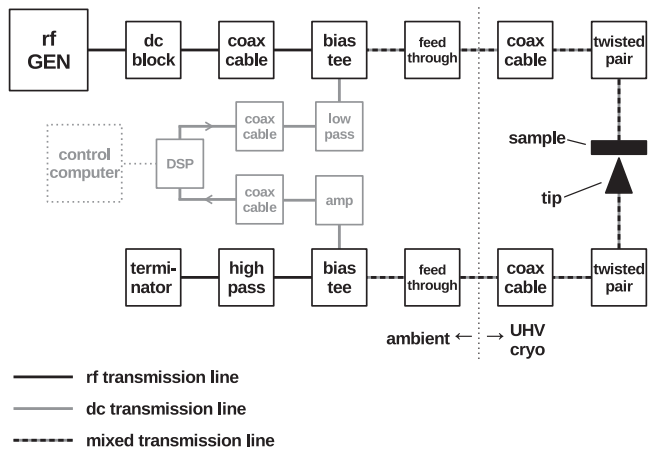

**FIG. 1.** Schematic of the electronic setup of our rf STM, highlighting the different components of the rf, dc, and mixed transmission lines. The control computer and the region of cryogenic UHV conditions are indicated with dotted lines. DSP stands for digital signal processor, rf GEN for rf signal generator, and amp for current amplifier.

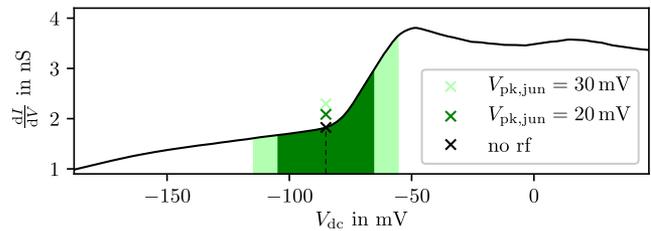

**FIG. 2.** d$I$/d$V$-spectrum of Ag(111) and the shift of one d$I$/d$V$-value when rf voltage is applied. The black solid curve is the average of 125 single d$I$/d$V$-spectra. The value of the d$I$/d$V$-spectrum at $V_{dc} = -85$ mV is marked with a black cross. When rf voltage is applied, the measured d$I$/d$V$-value increases due to rectification at the step, as explained in the text. The dark green cross marks the value when the rf voltage amplitude is 20 mV and the light green cross when it is 30 mV. The shaded areas of the same colors illustrate the intervals that influence the corresponding d$I$/d$V$-value at $V_{dc} = -85$ mV when rf voltage is applied.

Here, $t$ is the time, $V_{dc}$ is the slowly time-varying component (≤few kHz), and $V_{rf}$ is the fast time-varying component (rf voltage, MHz–GHz) with an amplitude of up to ≈100 mV, limited by the attenuation in the transmission line.

It is convenient to consider the electronic setup as three parts as shown in Fig. 1. The dc and rf parts cover the electronics and cabling associated purely with the STM control and frequency modulation, respectively. In the mixed part, from the bias tees onward, the signals are transmitted through the same cables.

Although the insertion loss of the commercial electrical components is well specified, the determination of the total insertion loss of the whole transmission line from the signal generator to the tunnel junction is not straightforward due to the complicated rf characteristics of the last few centimeters of the line, including the sample holder and sample.[11] A practical measurement of the total insertion loss is based on determining $V_{pk,jun}$.[11,13,14,17] We determine $V_{pk,jun}$ with the procedure reported by Paul et al.[14]

If the power at the rf signal generator output ($P_{gen}$) is held constant, $V_{pk,jun}$ varies with changing frequency. Therefore, it is necessary to adjust $P_{gen}$ to achieve a constant, frequency-independent $V_{pk,jun}$. We call this adjustment calibration of $V_{pk,jun}$.

For the calibration method, we make use of the rectification effect of the nonlinearity (step) in the d$I$/d$V$-spectrum of Ag(111). Rectification of the tunneling current causes a broadening of the step in the d$I$/d$V$-spectrum if a rf voltage is applied.[14,17] Consequently, in the vicinity of the nonlinearity, the magnitude of the differential tunneling conductance depends on $V_{pk,jun}$. This dependence is illustrated in Fig. 2 for a $V_{dc}$-value at the bottom of the step; the d$I$/d$V$-value increases with $V_{pk,jun}$ and thus with $P_{gen}$.

$V_{pk,jun}$ is frequency-independent if the d$I$/d$V$-value measured at a fixed $V_{dc}$ close to the nonlinearity is the same for each frequency. For Ag(111), a $V_{dc}$ of −85 mV is chosen because it is near the lower edge of the step (see Fig. 2). For each frequency, $P_{gen}$ is adjusted until the d$I$/d$V$-value reaches a predefined target (see Fig. 3).

One advantage of this method is that it is not necessary to know the exact functional relation between $P_{gen}$ and the d$I$/d$V$-value. In particular, the relation does not need to be linearized nor fitted, unlike in previously reported methods.[14,17] The relation has to be bijective in the region of interest, which can be verified by recording d$I$/d$V$ at a fixed $V_{dc}$ while sweeping $P_{gen}$. Due to the limited measurement time and the possibility to choose a low rf power, the influences from tip motion and rf-induced heating can be kept to a minimum, although they have to be monitored. Systematic errors introduced by the feedback loop that adjusts $P_{gen}$ have to be carefully considered. The nonlinearity causing the rectification has to be independent of the frequency and independent of the rf power.

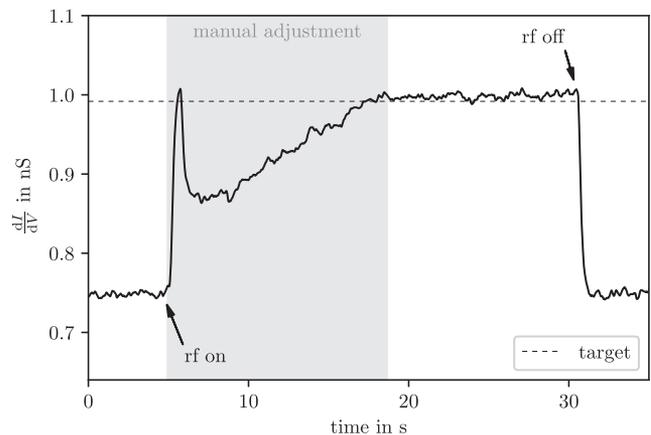

**FIG. 3.** Exemplary time evolution of the d$I$/d$V$-value at constant $V_{dc}$ during calibration. In the first and last ≈4 s, no rf voltage was applied. The times when the rf signal generator was switched on and off are marked with the annotations *rf on* and *rf off*, respectively. After the rf signal generator was switched on, $P_{gen}$ was adjusted until the target value was reached. The adjustment period is marked as *manual adjustment* and shaded in gray. The target d$I$/d$V$-value is drawn as a horizontal line. This example also highlights some potential issues with the manual procedure. Namely, the peak at ≈6 s is caused by an unintentional fast variation of $P_{gen}$. The slower increase after ≈7 s reflects the typical pace. In addition, it is clear that some inaccuracy in the determination of the target value remains as the d$I$/d$V$-value between ≈20 and ≈30 s is slightly too high. Automation of the procedure will improve this.

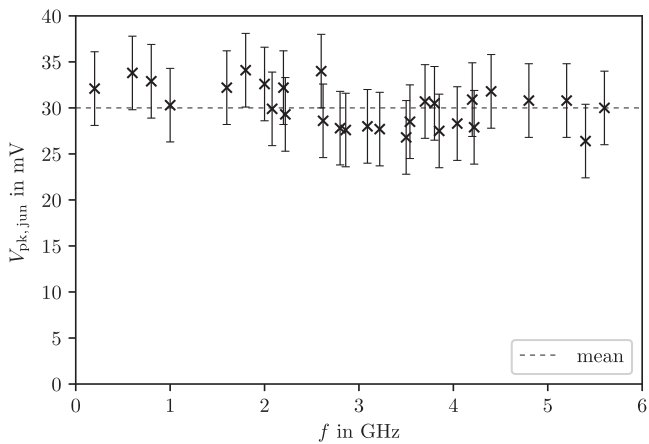

**FIG. 4.** Radio-frequency voltage amplitudes for 29 frequency values up to 5.6 GHz after manual calibration to the target value of 30 mV. The mean value of the calibrated amplitudes is (30 ± 2) mV, depicted by the dashed line. The frequency values were chosen by relevance to another experiment and are therefore not evenly distributed. With the future automation of the calibration, we intend to reduce the variance of the amplitudes. Improvements of the cabling and instrumentation will reduce the measurement error.

The success of the calibration can be quantified by determining the calibrated $V_{\mathrm{pk,jun}}$ at each frequency and comparing it to the desired value. Figure 4 shows the result of a manual calibration. The desired value of 30 mV is met by the calibrated rf voltage amplitudes, which have a mean value of (30 ± 2) mV. With future automation, we expect to reduce the variance of the calibrated amplitudes. The calibration for each frequency value takes about two minutes, and the measurement of $V_{\mathrm{pk,jun}}$ takes about 10 min, including analysis and documentation. Once the automation is implemented, we expect both procedures to become significantly faster.

Radio-frequency scanning tunneling spectroscopy is a powerful tool enabling, e.g., single-atom spin resonance. It requires a frequency-independent rf voltage amplitude across the tunnel junction, which is nontrivial. We have developed a method to calibrate the rf voltage amplitude to a frequency-independent value that is complementary to published methods with certain differences and advantages. It is advantageous when a quick rf voltage amplitude equalization for a small number of frequency values is needed as well as when changes in the rf transfer function occur frequently. The first results show the successful application of our method: after calibration to the target value of 30 mV, we measure a mean rf peak voltage across the tunnel junction of (30 ± 2) mV at 29 different frequencies up to 5.6 GHz. Although this work has been achieved with a rf STM, the described procedure is applicable to any other device that can deliver rf voltage to a tunnel junction.

This project received funding from the European Research Council (ERC) under the European Union's Horizon 2020 research and innovation programme (Grant Agreement No. 771193). The authors acknowledge the financial support from the Government of the Province of Upper Austria together with the Johannes Kepler University Linz (LIT Project No. 2016-1-ADV-002).

## DATA AVAILABILITY

The data that support the findings of this study are available from the corresponding author upon reasonable request.